\newif\ifanonymous
\documentclass[sigconf]{acmart}

\usepackage{todonotes}
\usepackage{hyperref}

\usepackage{listings, xcolor}

\definecolor{verylightgray}{rgb}{.97,.97,.97}

\lstdefinelanguage{Solidity}{
	keywords=[1]{anonymous, assembly, assert, balance, break, call, callcode, case, catch, class, constant, continue, constructor, contract, debugger, default, delegatecall, delete, do, else, emit, event, experimental, export, external, false, finally, for, function, gas, if, implements, import, in, indexed, instanceof, interface, internal, is, length, library, log0, log1, log2, log3, log4, memory, modifier, new, payable, pragma, private, protected, public, pure, push, require, return, returns, revert, selfdestruct, send, solidity, storage, struct, suicide, super, switch, then, this, throw, transfer, true, try, typeof, using, value, view, while, with, addmod, ecrecover, keccak256, mulmod, ripemd160, sha256, sha3}, % generic keywords including crypto operations
	keywordstyle=[1]\color{blue}\bfseries,
	keywords=[2]{address, bool, byte, bytes, bytes1, bytes2, bytes3, bytes4, bytes5, bytes6, bytes7, bytes8, bytes9, bytes10, bytes11, bytes12, bytes13, bytes14, bytes15, bytes16, bytes17, bytes18, bytes19, bytes20, bytes21, bytes22, bytes23, bytes24, bytes25, bytes26, bytes27, bytes28, bytes29, bytes30, bytes31, bytes32, enum, int, int8, int16, int24, int32, int40, int48, int56, int64, int72, int80, int88, int96, int104, int112, int120, int128, int136, int144, int152, int160, int168, int176, int184, int192, int200, int208, int216, int224, int232, int240, int248, int256, mapping, string, uint, uint8, uint16, uint24, uint32, uint40, uint48, uint56, uint64, uint72, uint80, uint88, uint96, uint104, uint112, uint120, uint128, uint136, uint144, uint152, uint160, uint168, uint176, uint184, uint192, uint200, uint208, uint216, uint224, uint232, uint240, uint248, uint256, var, void, ether, finney, szabo, wei, days, hours, minutes, seconds, weeks, years},	% types; money and time units
	keywordstyle=[2]\color{teal}\bfseries,
	keywords=[3]{block, blockhash, coinbase, difficulty, gaslimit, number, timestamp, msg, data, gas, sender, sig, value, now, tx, gasprice, origin},	% environment variables
	keywordstyle=[3]\color{violet}\bfseries,
	identifierstyle=\color{black},
	sensitive=false,
	comment=[l]{//},
	morecomment=[s]{/*}{*/},
	commentstyle=\color{gray}\ttfamily,
	stringstyle=\color{red}\ttfamily,
	morestring=[b]',
	morestring=[b]"
}

\AtBeginDocument{%
  \providecommand\BibTeX{{%
    \normalfont B\kern-0.5em{\scshape i\kern-0.25em b}\kern-0.8em\TeX}}}

\setcopyright{none}
\renewcommand\footnotetextcopyrightpermission[1]{} % removes footnote with conference information in first column
\newcommand{\ParGen}{\mathsf{ParGen}}
\newcommand{\KGen}{\mathsf{KGen}}
\newcommand{\UKGen}{{\user.\KGen}}
\newcommand{\DKGen}{{\delegate.\KGen}}
\newcommand{\Sign}{\mathsf{Sign}}
\newcommand{\USign}{{\user.\Sign}}
\newcommand{\DSign}{{\delegate.\Sign}}
\newcommand{\Verify}{\mathsf{Verify}}

\newcommand{\BCUpdate}{\mathsf{BCUpdate}}
\newcommand{\Delegate}{\mathsf{Delegate}}

\newcommand{\SigParGen}{\mathsf{SParGen}}
\newcommand{\SigKGen}{\mathsf{SKGen}}
\newcommand{\SigSign}{\mathsf{SSign}}
\newcommand{\SigVerify}{\mathsf{SVerify}}

\newcommand{\Enc}{\mathsf{Enc}}

\newcommand{\ZKCRS}{\mathsf{ZKSetup}}
\newcommand{\ZKProof}{\mathsf{ZKProof}}
\newcommand{\ZKVerify}{\mathsf{ZKVerify}}

\newcommand{\statement}{y}
\newcommand{\witness}{w}
\newcommand{\relation}{R}

\newcommand{\sigscheme}{\Sigma}
\newcommand{\zkscheme}{\Pi}

\newcommand{\sk}{\mathsf{sk}}
\newcommand{\pk}{\mathsf{pk}}

\newcommand{\usk}{\mathsf{usk}}
\newcommand{\upk}{\mathsf{upk}}
\newcommand{\dsk}{\mathsf{dsk}}
\newcommand{\dpk}{\mathsf{dpk}}
\newcommand{\esk}{\mathsf{osk}}

\newcommand{\GG}{\mathcal{G}}
\newcommand{\ZZ}{\mathbb{Z}}

\newcommand{\sig}{\sigma}
\newcommand{\contract}{sc}
\newcommand{\trigger}{tr}
\newcommand{\st}{st}
\newcommand{\aux}{aux}

\newcommand{\NIZK}{\mathsf{NIZK}}
\newcommand{\hash}{\mathsf{h}}

\newcommand{\delegate}{\mathcal{D}}
\newcommand{\user}{\mathcal{U}}

\newcommand{\setS}{\mathcal{S}}
\newcommand{\algA}{\mathsf{A}}

\newcommand\getsr{\mathrel{%
    \mathchoice{\qmf}{\qmf}{\scriptsize\qmf}{\tiny\qmf}%
}}
\newcommand\qmf{{%
    \setbox0\hbox{$\gets$}%
    \rlap{\hbox to \wd0{\hss\raisebox{1.2\height}{\tiny\$}\hss}}\box0
}}

\newcommand{\pp}{pp}
\newcommand{\crs}{crs}
\newcommand{\secpar}{\lambda}
\newcommand{\msg}{m}

\begin{document}

\title{Single-Use Delegatable Signatures Based on Smart Contracts}

\ifanonymous
	\author{Anonymous submission}
	%\orcid{0000}
	\email{}
	\affiliation{%
	  \institution{~}
	  \streetaddress{~}
	  \city{~}
	  \country{}
	  \postcode{}
	}

	\renewcommand{\shortauthors}{Anonymous submission to ARES 2021}
\else
	\author{Stephan Krenn}
	%\orcid{0000}
	\email{stephan.krenn@ait.ac.at}
	\affiliation{%
	  \institution{AIT Austrian Institute of Technology}
	  \streetaddress{Giefinggasse 4}
	  \city{Vienna}
	  \country{Austria}
	  \postcode{1201}
	}

	\author{Thomas Lor\"unser}
	%\orcid{0000}
	\email{thomas.loruenser@ait.ac.at}
	\affiliation{%
	  \institution{AIT Austrian Institute of Technology}
	  \streetaddress{Giefinggasse 4}
	  \city{Vienna}
	  \country{Austria}
	  \postcode{1201}
	}

\fi


\begin{abstract}
  Delegation of cryptographic signing rights has found many application in the literature and the real world.
  However, despite very advanced functionalities and specific use cases, existing solutions share the natural limitation that the number of usages of these signing rights cannot be efficiently limited, but users can at most be disincentivized to abuse their rights.

  In this paper, we suggest a solution to this problem based on blockchains.
  We let a user define a smart contract defining delegated signing rights, which needs to be triggered to successfully sign a message.
  By leveraging the immutability of the blockchain, our construction can now guarantee that a user-defined threshold of signature invocations cannot be exceeded, thereby circumventing the need for dedicated hardware or similar assistance in existing constructions for one-time programs.
  
  We discuss different constructions supporting different features, and provide concrete implementations in the Solidity language of the Ethereum blockchain, proving the real-world efficiency and feasibility of our construction.
\end{abstract}


%\ccsdesc[500]{Security and privacy~Digital signatures}
%\ccsdesc[300]{Security and privacy~Distributed systems security}


\maketitle
\pagestyle{plain}

\newcommand\blfootnote[1]{%
  \begingroup
  \renewcommand\thefootnote{}\footnote{#1}%
  \addtocounter{footnote}{-1}%
  \endgroup
}

\blfootnote{This is the author's version of the work. It is posted here for your personal use. Not for redistribution. The definitive Version of Record was published in \textit{The $16$th International Workshop on Frontiers in Availability, Reliability and Security (FARES 2021)}, DOI \href{https://doi.org/10.1145/3465481.3469192}{10.1145/3465481.3469192}.}

\section{Introduction}
Digital signatures are the central cryptographic primitive to provide strong and provable authenticity and integrity guarantees.
Over the last decades, numerous advanced, so-called malleable, signature schemes have been introduced, which allow the holder of the secret key to delegate certain signature rights to a delegate.
Examples for such signature schemes include proxy signatures~\cite{JC:BolPalWar12}, poly-based signatures~\cite{PKC:BelFuc14}, functional signatures~\cite{PKC:BoyGolIva14}, blank signatures~\cite{DBLP:journals/ijisec/WangPD18}, redactable signatures~\cite{ICISC:SteBulZhe01}, sanitizable signatures~\cite{PKC:CDKPSS17}, or protean signatures~\cite{CANS:KPSS18};
for a detailed overview, we refer to Bilzhause et al.~\cite{DBLP:conf/IEEEares/BilzhausePS17}.

Despite significant differences in terms of functionality and addressed use cases, all these schemes share the natural limitation that delegated rights are not limited in the number of usages of these delegated rights.
However, this might be a very desirable property in many applications, e.g., for blank cheques.

\paragraph{Related work}

In the following we give an overview of diverse existing approaches that aim at enforcing a single usage of cryptographic keys, which may also be adapted for the case of delegated signature rights.

One-time programs (OTPs), introduced by Goldwasser et al.~\cite{C:GolKalRot08}, are computer programs that can only be executed a single time, and self-destruct afterwards.
However, it is easy to see that such programs cannot be fully software-based, as it is always possible to copy and re-execute a piece of software.
Consequently, constructions for one-time programs found in the literature require tamper-proof hardware, e.g.,~\cite{C:GolKalRot08,TCC:GISVW10,DBLP:conf/cisc/DurnogaDKZ13} or are based on trusted execution environments~\cite{FC:ZCDBMAC19}.

While it is known that perfect, information theoretically secure one-time programs are not even possible in the quantum setting~\cite{C:BroGutSte13}, Roehsner et al.~\cite{roehsner18} proposed a probabilistic quantum-based solution for OTPs.
In a nutshell, the idea is to encode the program onto quantum states in a way that it needs to be measured for evaluation, in which case the system changes its state and can no longer be reused.
Furthermore, by the no-cloning theorem, the quantum program cannot be copied before execution.
Specifically for the case of delegated signing rights, Roehsner et al.~\cite{roehsner18} showed how to make the failure probability arbitrarily small.
However, even though the resulting signature is classical, their construction is not (yet) of practical interest as it requires to transfer quantum states between the user and the delegate, and to maintain this quantum state until measurement. 

A complementary approach to prevent double-spending of cryptographic tokens (e.g., signature keys, e-cash, etc.) is to guarantee that some secret key is revealed if two linked activities are performed, e.g.,~\cite{SCN:CatFucSol20,ESORICS:PoeSte14,PKC:BelPoeSte17,ASIACCS:BEKSS20,PKC:BCFK15}.
While this approach disincentivizes users to use cryptographic tokens multiple times, its practical usability is sometimes limited.
Depending on the application scenario it might be difficult to actually detect such a double usage.
Furthermore, in this case, also potential physical consequences of issued signatures need to be annulled, causing substantial overhead.
Specifically in the context of e-cash, online double-spending prevention avoids this problem by constantly contacting the issuing bank to detect whether a specific coin has already been spent~\cite{C:Chaum83a}.
This approach is well suited for centralized systems, but does not scale for scenarios where, e.g., single signature rights are to be delegated without the existence of a central party.

Finally, recently, Goyal and Goyal~\cite{TCC:GoyGoy17} proposed a generic solution for OTPs leveraging proof-of-stake based blockchains, without requiring trusted setup or random oracles.

\paragraph{Our contribution.}
  In this paper, we propose a simple yet elegant solution which allows one to enforce a single use of a cryptographic token.
  Our solution is based on smart contracts, and intuitively works as follows:
  when delegating signing rights, the owner of a secret key publishes a smart contract, which may later be triggered by the delegate holding a secret key.
  By executing the smart contract, its state changes such that it cannot be invoked any more.
  Furthermore, the state contains a blinded hash of the signed message, such that the verifier can check that the received signature was indeed the first signature issued by this delegate.
  
  We propose multiple variants for our mechanism for different scenarios, e.g., depending on whether a dedicated delegate is required or whether transparency is important.
  
  Finally, we provide concrete implementations of our schemes using Ethereum's Solidity specificaton language for smart contracts, and demonstrate the real-world efficiency and cheapness of our solution.

\section{Preliminaries}
We will next introduce the notation and required background that will be used in the remainder of this document.

\subsection{Notation}
  We denote the main security parameter by $\secpar$.
  For a finite set $\setS$, we write $s\getsr\setS$ to denote that $s$ was sampled uniformly at random in $\setS$.
  Similarly, we write $a\getsr\algA(b)$ to denote that $a$ is assigned the outputs of a potentially randomized algorithm $\algA$ on input $b$.
  All algorithms discussed throughout this paper are probabilistic polynomial time.

\subsection{Cryptographic Background}
We briefly recap the notation and security properties for digital signatures and zero-knowledge proofs, as well as our modeling of blockchains, but omit details due to space limitations and refer to the original literature.

\paragraph{Digital signatures.}
  A digital signature scheme consists of four algorithms $(\SigParGen,\SigKGen,\SigSign,\SigVerify)$:
  \begin{description}
    \item[$\pp_\sigscheme\getsr\SigParGen(1^\secpar)$.]
      On input the security parameter, this algorithm outputs the public parameters $\pp_\Sigma$, which are assumed to be implicit input to all further algorithms.
    \item[$(\sk,\pk)\getsr\SigKGen(\pp_\sigscheme)$.]
      On input the public parameters, this algorithm outputs a secret signing key $\sk$ and a corresponding verification key $\pk$.
    \item[$\sig\getsr\SigSign(\sk,\msg)$.]
      On input a secret signing key and a message, this algorithm outputs a signature $\sig$.
    \item[$b\gets\SigVerify(\pk,\msg,\sig)$.]
      On input a verification key, a message, and a signature, this algorithm outputs a bit indicating whether to accept or to reject the signature.
  \end{description}
  Informally, a signature scheme is correct if every honestly generated signature also passes the verification algorithm.
  Furthermore, the scheme is said to be existentially unforgeable under chosen-message attacks (EUF-CMA), if no adversary can produce a valid on a new signature, even after having seen arbitrarily many signatures on messages of his choice.
  For a formal discussion we refer to the literature~\cite{GolMicRiv88}.
  
\paragraph{Zero-knowledge proofs.}
  Non-interactive zero-knowledge proof of knowledge consists of three algorithms $(\ZKCRS,\allowbreak \ZKProof,\allowbreak \ZKVerify)$:
  \begin{description}
    \item[$\crs_\zkscheme\getsr\ZKCRS(1^\secpar)$.]
      On input the security parameter, this algorithm outputs the common reference string $\crs_\zkscheme$ for the proof system, which is assumed to be implicit input to all further algorithms.
    \item[$\pi\getsr\ZKProof(\statement,\witness)$.]
      On input a statement $\statement$ and a corresponding $\witness$ such that $(\statement,\witness)\in\relation$, this algorithm outptus a non-interactive zero-knowledge proof of knowledge for $\witness$.
    \item[$b\gets\ZKVerify(\statement,\pi)$.]
      On input a statement $\statement$ and a proof $\pi$, this algorithm outputs a bit indicating whether to accept or to reject the proof.
  \end{description}
  Intuitively, such a proof system needs to be correct, i.e., every honestly generated proof should also pass the verification algorithm.
  The zero-knowledge property requires that no adversary can infer any information about $\witness$ only knowing $\crs_\zkscheme$, $\statement$ and $\pi$;
  this is modeled through a simulator knowing a trapdoor to a (simulated) $\crs_\zkscheme$ which, given as input a statement, generates simulated proofs that are indistinguishable from honestly generated ones.
  Finally, extractability requires that an adversary not knowing a valid witness is incapable of generated a valid proof for a given $\statement$;
  again, this is modeled through the existence of an algorithm, which, knowing a trapdoor to a (simulated) $\crs_\zkscheme$ can efficiently extract a valid $\witness$ for every accepting proof $\pi$ for a given statement.
  For formal definitions and further discussion, we refer, e.g., to~\cite{CSF:CKLM14}.

  For readability, we will use the notation introduced by Camenisch and Stadler~\cite{C:CamSta97} to denote zero-knowledge proofs.
  That is, we will write:
  $$
    \NIZK\left[(x_1,x_2):Y_1=x_1G \quad\land\quad Y_2=x_2H\right](\msg)
  $$
  to denote a non-interactive zero-knowledge proof of knowledge of $x_1,x_2$ such that the relation on the right hand side is satisfied.
  All protocols used in this paper can efficiently be instantiated in the random-oracle model using $\Sigma$-protocols~\cite{C:Schnorr89,cramer97} and the Fiat-Shamir heuristic~\cite{C:FiaSha86}, which will also be used to bind a proof to a given $\msg$.
  
  Following the observations of Bernhard et al~\cite{AC:BerPerWar12}, we assume that all relevant context--and in particular the statement to be proven--is used when computing the challenge in the Fiat-Shamir transform, even though we do not make this explicit to not disguise the notation.

\paragraph{Blockchains.}
  In this paper, we consider a blockchain as a permissionless, public bulletin board with  two natural properties.
  Namely, we require immutability, meaning that information written to the blockchain can not be altered or deleted, and we assume that adversarial forks can efficiently be distinguished from the actual block chain state.
  For detailed discussions, we refer, e.g., to Goyal and Goyal~\cite{TCC:GoyGoy17}.

\section{Definitions}
The following sections first introduce the syntax and notation for single-use delegatable signatures, and then summarize the security requirements posed to such schemes.

\subsection{Syntax}
A single-use delegatable signature scheme consists of the following set of algorithms:
\begin{description}
  \item[$\pp\getsr\ParGen(1^\secpar)$.] 
    On input the security parameter, this algorithm outputs public parameters $\pp$.
  \item[$(\usk,\upk)\getsr\UKGen(\pp)$.] 
    On input the public parameters, this algorithm outputs a secret key $\usk$ and a corresponding public key $\upk$ for a user.
  \item[$(\dsk,\dpk)\getsr\DKGen(\pp)$.] 
    On input the public parameters, this algorithm outputs a secret key $\dsk$ and a corresponding public key $\dpk$ for a delegate.
  \item[$(\esk,\contract)\getsr\Delegate(\usk,\dpk,\aux)$.] 
    On input a user's secret key, a delegate's public key, and some auxiliary information $\aux$ this algorithm outputs a delegated one-time signature key $\esk$, as well as a value $\contract$ (which in our case will be a smart contract published in a blockchain).
  \item[$(\sig,\trigger)\getsr\DSign(\dsk,\esk,\upk,\msg)$.] 
    This algorithm allows a delegate holding a one-time key to compute a signature $\sig$ on a message $\msg$.
    Furthermore, the algorithm outputs an auxiliary value $\trigger$ (which in our case will be trigger for the smart contract).
  \item[$(\sig,\trigger)\getsr\USign(\usk,\esk,\dpk,\msg)$.] 
    This algorithm allows the user to compute a signature on a message as well as an auxiliary value $\trigger$.
  \item[$\contract'\getsr\BCUpdate(\contract,\trigger,\aux)$.]
    Knowing $\trigger$, this algorithm updates the state $\contract$ (which in our case will be a modification of the state of smart contract).
  \item[$b\gets\Verify(\upk,\dsk,\msg,\sig,\contract,\aux)$.] 
    This message outputs a bit indicating whether to accept or to reject a signature for a given message depending also on $\contract$ and $\aux$ (which in our case will be the state of the blockchain).
\end{description}

\subsection{Security Requirements}
In the following we informally discuss the security requirements expected from a single-use delegatable signature scheme.
A full formalization of these requirements is left for future work.

\paragraph{Completeness.}
Completeness requires that, if all parties behave honestly, signatures will always verify correctly.

\paragraph{Unforgeability.} 
Strong unforgeability requires that an adversary neither knowing the user's nor the delegate's secret key and the one-time key can generate a valid signature on its own.
In the case that an adversary can generate new signatures on messages previously signed by the user or the delegate, we say that the scheme satisfies weak unforgeability.

\paragraph{Transparency.}
For malleable signatures, transparency typically requires that an outsider not knowing any secret keys can decide whether a valid signature has been generated by the user or by the delegate.
In the context of our work we additionally require that also the originator of $\trigger$ cannot be determined, in order to also guarantee transparency during the $\BCUpdate$ process.

\paragraph{Onetimeness.}
Onetimeness requires that a delegated signature key $\esk$ can only be used to sign a single message, either by the user or by the delegate.
Any further attempt to re-use $\esk$, even by a legitimate user, will result in an invalid signature.
As for unforgeability, we distinguish between weak and strong onetimeness, depending on whether multiple valid signatures for the same message can be generated or not.  

\section{Constructions}
In the following we present constructions of single-use delegatable signatures.
We first present a very basic scheme where delegated rights can be forwarded to third party (yet only consumed once).
We then present a scheme with a designated verifier, and subsequently discuss possible extensions to achieve accountability, $n$-time signatures, and more.

\subsection{A Basic Scheme}\label{sec:basic}
The idea of our basic scheme is that the user puts a signed commitment by means of a smart contract into a blockchain, and gives the opening of the commitment to a delegate.
To sign, the delegate provides a zero-knowledge proof of knowledge of the opening to the blockchain network, which verifies the proof and locks the smart contract by storing the hash of the signed message.

Somewhat surprisingly, for the basic construction, the delegate does not need to own any local secret key, i.e., $\dsk$ can be set to $\bot$;
furthermore, when signing a message using the delegated signing key, the delegate does not need to generate an actual signature (i.e., $\sigma=\bot$):
because of the soundness of the NIZK, already the fact that $\hash(\msg)$ is stored in the smart contract suffices to convince the verifier that a legitimate entity (i.e., the signer or the delegate) have triggered the signing process.

\medskip

In the following presentation, let $(\SigKGen,\SigSign,\SigVerify)$ be a EUF-CMA secure signature scheme.

\begin{itemize}
\item
  $\ParGen(1^\secpar)$ outputs $\pp=(1^\secpar,\GG,G,q)$, where $\GG=\langle G\rangle$ is a cyclic group of prime order $q$, such that the discrete logarithm problem is hard in $\GG$.

\item
  $\UKGen(\pp)$ outputs a key pair $(\usk,\upk)\getsr\SigKGen(1^\secpar)$.

\item
  $\DKGen(\pp)$ outputs $(\dsk,\dpk)=(\bot,\bot)$.

\item
  $\Delegate(\usk,\dpk,\aux)$ samples $\esk\getsr\ZZ_q$.
The algorithm furthermore computes $Y=\esk\cdot G$ and $\tau=\SigSign(\usk,(Y,\aux))$, where $\aux$ is a unique identifier of the block into which the smart contract will be inserted in the blockchain.
Finally, the algorithm defines $\contract$ as a stateful smart contract for the following functionality:
\begin{itemize}
  \item
    The contract fixes $\pp$, $Y$, $\tau$, and $\upk$, and initializes its internal state as $\st=\varepsilon$.
  \item
    Being called on input $\pi$ and $\hash$, the contract first checks whether $\st=\varepsilon$ and aborts if this is not the case.
  \item
    It then checks whether $\pi$ is a valid NIZK of $\esk$ such that $Y=\esk\cdot G$.
    If this is the case, it sets $\st=\hash$.
\end{itemize}

\item
  $\DSign(\dsk,\esk,\upk,\msg)$ computes $\hash=\hash(\msg)$ and 
  $$
  \pi\gets\NIZK[(\esk):Y=\esk\cdot G](\hash)\,.
  $$
It outputs $\sig=\bot$ and $\trigger=(\pi,\hash)$.

\item
  $\USign(\usk,\esk,\dpk,\msg)$ computes $\hash=\hash(\msg)$ and 
  $$
  \pi\gets\NIZK[(\esk):Y=\esk\cdot G](\hash)\,.
  $$
It outputs $\sig=\bot$ and $\trigger=(\pi,\hash)$.

\item
  $\BCUpdate(\contract,\trigger)$ checks that $\trigger=(\pi,\hash)$ is as defined in the $\contract$ and ouputs $\contract$ with the potentially updated $\st$.

\item
  $\Verify(\upk,\dsk,\msg,\sigma,\contract)$ outputs $1$ if and only if $\contract$ contains a valid signature for $\upk$ and if $\st=\hash(\msg)$.
\end{itemize}

\subsubsection{Security Considerations}
Correctness of the scheme follows immediately by inspection.

Regarding unforgeability, one can see that the smart contract $\sc$ is bound to the specific block on the blockchain by the inclusion of the block identifier in the user's signature $\tau$.
Therefore, any attempt to forge a signature would need to either forge a signature of the underlying EUF-CMA signature scheme, or leverage the given instance of the smart contract $\contract$.
Now, by the soundness properties of the deployed NIZK system, it follows that knowledge of $\esk$ is required to generate a valid $\trigger$ to activate the execution of $\contract$.

From the immutability properties and the usual soundness assumptions of the block chain (honest majority, etc.), it furthermore follows that the smart contract can only be executed once, and thus strong onetimeness follows.

Finally, the distributions of signatures generated by the user and the delegate are identical, and thus transparency follows immediately.

Formal proofs, together with formal security definitions, are planned for future work.

\subsubsection{Discussion}
The above construction does not define a dedicated delegate.
That is, the delegate could further delegate the signing rights by simply forwarding $\esk$ to a third party, without having to reveal any sensitive private key material.
While this may be desirable in certain situations, we will discuss in the following constructions where the delegate is defined by the user and forwarding of signing rights is prohibited.
However, it is worth noting that this delegation does not contradict our unforgeability definition, as the $\dsk=\bot$ would be known to a third party, and therefore the requirement that $\dsk$ and $\esk$ need to be known to generate a signature would be satisfied.
Furthermore, we note that also onetimeness is not affected by forwarding $\esk$, as the blockchain network would still only accept the first invocation of $\contract$.  

We also note that somewhat surprisingly it is not necessary to attach an actual signature to the signed message, as the pure fact that a hash value is stored in the smart contract's state is sufficient to prove the authenticity of the message.
However, certain situations might require that for privacy reasons this hash value does not enable an attacker to reconstruct the signed message.
This can be addressed by simply replacing $\hash=\hash(\msg)$ by $\hash=\hash(r,\msg)$ for $r\gets\{0,1\}^{2\secpar}$, and defining $\sigma=r$ as the actual signature.
By doing so, it is guaranteed that $\hash$ statistically hides any information about $\msg$. 

Finally, the value of $\esk=x$ does not need to be kept secret any longer once $\contract$ has been updated and the signing right has been consumed.
However, the NIZK cannot simply be replaced by sending $x$ in the plain, as malicious nodes could otherwise modify $\hash$ without being detected before the transaction has been sufficiently distributed within the network.
It is thus important to bind the knowledge of $x$ to the value of $\hash$.

\subsection{Adding a Designated Delegate}\label{sec:designated}
We next present an extension of the basic scheme which allows for a dedicated designate.
To achieve this, it becomes necessary for delegates to have sensitive private keys.

For a basic scheme, the user would now simply sign the delegate's public key $\dpk$ instead of $Y$ in the $\Delegate$ algorithm.
In order to trigger the smart contract, the delegate would then no longer compute a NIZK for the discrete logarithm of $Y$, but for $\dsk$ corresponding to $\dpk$.
In order to also give the user the option to trigger the smart contract herself, also a NIZK for $\usk$ corresponding to $\upk$ would be accepted.
It would now be guaranteed that only the holder of $\dsk$ or of $\usk$ could sign the message, and thus the delegate could no longer forward signing rights as this would require to reveal $\dsk$.

However, while this construction is complete, unforgeable, and onetime, it does not achieve transparency, as the statement proven by the NIZK would reveal whether it was generated by the user or the delegate.
In order to achieve symmetry, the NIZK thus shows that one either knows the user's or the delegate's secret key.

\medskip

More precisely, the full construction for a designated-delegate signature scheme is given by the following algorithms:

\begin{itemize}
\item
  $\ParGen(1^\secpar)$ outputs $\pp=(1^\secpar,\GG,G,q)$, where $\GG=\langle G\rangle$ is a cyclic group of prime order $q$, such that the discrete logarithm problem is hard in $\GG$.

\item
  $\UKGen(\pp)$ computes a key pair $(\usk',\upk')\getsr\SigKGen(1^\secpar)$.
  Furthermore, it chooses $\usk''\getsr\ZZ_q$ and sets $\upk''\gets\usk''\cdot G$.
  Finally, it outputs 
  $$(\usk,\upk)=((\usk',\usk''),(\upk',\upk'')).$$

\item
  $\DKGen(\pp)$ chooses $\dsk\getsr\ZZ_q$ and sets $\dpk\gets\dsk G$.

\item
  $\Delegate(\usk,\dpk,\aux)$ sets $\esk=\bot$.
  It then parses $\usk=(\usk',\usk'')$ and computes $\tau=\SigSign(\usk',(\dpk,\aux))$.
  Here $\aux$ is a unique identifier of the block into which the smart contract will be inserted in the blockchain.
Finally, the algorithm defines $\contract$ as a stateful smart contract for the following functionality:
\begin{itemize}
  \item
    The contract fixes $\pp$, $\dpk$, $\tau$, and $\upk=(\upk',\upk'')$, and initializes its state $\st=\varepsilon$.
  \item
    Being called on input $\pi$ and $\hash$, the contract first checks whether $\st=\varepsilon$ and aborts if this is not the case.
  \item
    It then checks whether $\pi$ is a valid NIZK for $\usk''$ corresponding to $\upk''$ or for $\dsk$ corresponding to $\dpk$.
    If this is the case, it sets $\st=\hash$.
\end{itemize}

\item
  $\DSign(\dsk,\esk,\upk,\msg)$ computes $\hash=\hash(\msg)$ and 
  \begin{align*}
  \pi\gets\NIZK[(\dsk,\usk''):\ &\dpk=\dsk\cdot G \quad \lor \\
                                &\upk''=\usk''\cdot G](\hash)\,,
  \end{align*}
thereby using $\dsk$ as the witness and proving the first literal of the clause.
It outputs $\sig=\bot$ and $\trigger=(\pi,\hash)$.

\item
  $\USign(\usk,\esk,\dpk,\msg)$ computes $\hash=\hash(\msg)$ and 
  \begin{align*}
  \pi\gets\NIZK[(\dsk,\usk''):\ &\dpk=\dsk\cdot G \quad \lor\\
                                &\upk''=\usk''\cdot G](\hash)\,,
  \end{align*}
thereby using $\usk''$ as the witness and proving the second literal of the clause.
It outputs $\sig=\bot$ and $\trigger=(\pi,\hash)$.

\item
  $\BCUpdate(\contract,\trigger)$ checks that $\trigger=(\pi,\hash)$ is as defined in the $\contract$ and ouputs $\contract$ with the potentially updated $\st$.

\item
  $\Verify(\upk,\dsk,\msg,\sig,\contract)$ outputs $1$ if and only if $\contract$ contains a valid signature for $\upk'$ and if $\st=\hash(\msg)$.
\end{itemize}

Note that the NIZKs computed by the delegate and the user are indistinguishable due to the zero-knowledge property.

\subsection{Further Extensions} 
  Our basic constructions can be extended in various directions, depending on the specific needs and requirements of the use case.
  
  \paragraph{Accountability.}
    The constructions presented above do not offer any possibility to identify the originator of a specific signature, as both the user as well as the delegate could equally trigger the smart contract.
    Accountability enables a predefined third party acting as a judge to identify the signer, see, e.g., Beck et al~\cite{ACISP:BCDKPS17}.
    One way to achieve this in our protocols would be to let the signer encrypt its public key, and later prove that the public key contained in the ciphertext is the key for which the corresponding secret key is known.
    That is, the NIZK would be changed to the following: 
  \begin{align*}
    \pi\gets\NIZK[&(\dsk,\usk'',r):\\
                  & \left( \dpk=\dsk\cdot G \land c=\Enc(\dpk;r) \right) \quad \lor\\
                  & \left( \upk''=\usk''\cdot G \land c=\Enc(\upk'';r) \right) ](\hash)\,,
  \end{align*}
  where $c$ is an encryption of the public key with randomness $r$.
  Here, one can think of the encryption scheme as the ElGamal crypto system~\cite{ElGamal85}.
  
  For this proof, the delegate would use $\dsk$ and $r$ in order to prove the first statement, while the user could use $\usk''$ and $r$ to prove the latter statement.
  It is important to note that the NIZK implicitly also proves that the \emph{same} value for which the discrete logarithm is known, is also encrypted within the ciphertext $c$, and by the soundness property it is thus infeasible to encrypt a different value in order to escape accountability.

  \paragraph{Immutability.}
    For instance in the scenario of sanitizable or blank signatures, the user may wish to fix certain parts of the message a delegate can sign.
    To achieve this, the user commits to the restrictions (e.g., in form of a message template, or as a circuit which outputs $1$ if and only if the signed message was valid) as part of the smart contract, and hands over the opening of the commitment to the delegate.
    Now, depending on the privacy requirements--whether or not the restrictions may be known to the verifier--the delegate either forwards the opening to the verifier as part of $\sig$, or computes a NIZK proving that the (known) signed message is indeed valid with respect to the (secret) restrictions;
    note however that the latter may be computationally expensive depending on the valid modifications. 
    
  \paragraph{Multiple delegates and $n$-time signatures.}
    Our constructions can directly be extended to multiple delegates, by letting the user defining a list of public keys that are allowed to act on behalf of him.
    Also, $n$-times signatures can be obtained by storing a list of up to $n$ hash values before denying further execution of the smart contract.
    Here, though causing some computational overhead and thus increasing the costs of the smart contract, the user could hide the upper bound $n$ from the public by only signing a commitment on it, and the delegate could prove that the number of preceding invocations is smaller than the number hidden in the commitment.

\section{Evaluation}
  In the following we provide implementations of the schemes specified above in the Solidity language for smart contracts on the Ethereum blockchain.
  Our implementation partially leverages existing elliptic curve implementations in Solidity~\cite{snarktest,heiswap}, and was implemented using the Remix Suite for Solidity smart contracts, which was also used to compute the cost estimates.
  The resulting code is given in Listings~\ref{lst:basic} and~\ref{lst:advanced}.

  In contrast to the abstract specification of our schemes it is not necessary to let the user sign the smart contract in the concrete implementation, as in Ethereum every transaction is anyways signed, and the smart contract thus points back to its sender.
  If, however, a binding to an existing public key outside of Ethereum is important, including a signature as in the construction would be a straightforward modification.
  
  Note that locally executed algorithms (i.e., for signing and verification) are not depicted here due to space limitations, and as they do not need to be included in the contract.

\begin{lstlisting}[float=t,frame=tb,label=lst:basic,caption=Basic scheme as defined in Section~\ref{sec:basic}]
pragma solidity ^0.4.14;
pragma experimental ABIEncoderV2;
import "./altbn128.sol";

// One time delegatable signatures
contract Otds
{
    // contract state
    address public owner;
    uint256 state;
    Curve.G1Point comm;

    // constructor initializing state and delegation
    constructor(Curve.G1Point _comm) public {
        owner = msg.sender;
        state = 0;
        comm = _comm;
    }

    // function called by delegatee to sign once
    function OtdsSign(uint256 c, uint256 r, uint256 hmessage)
        public
    {
        require(state == 0, "Already signed.");
        Curve.G1Point memory tp = Curve.g1add(
                Curve.g1mul(Curve.P1(), r % Curve.N()),
                Curve.g1mul(comm, c % Curve.N()));
        uint256 cp =  uint256(
                keccak256(abi.encodePacked(
                comm.X, comm.Y, tp.X, tp.Y, hmessage)));
        require(c == cp, "Invalid proof.");
        state = hmessage;
    }
}
\end{lstlisting}
\begin{lstlisting}[float=t,frame=tb,label=lst:advanced,caption=Advanced scheme as defined in Section~\ref{sec:designated}]
pragma solidity ^0.4.14;
pragma experimental ABIEncoderV2;
import "./altbn128.sol";

// One time delegatable signatures
contract OtdsOR
{
    // contract state and public parameters
    address public owner;
    uint256 public state;
    Curve.G1Point g1;
    Curve.G1Point g2;
    Curve.G1Point y1;
    Curve.G1Point y2;
   
    constructor(Curve.G1Point _g1, Curve.G1Point _g2, 
        Curve.G1Point _y1, Curve.G1Point _y2)
        public
    {
        // init state and public parameters
        owner = msg.sender;
        state = 0;
        g1 = _g1;
        g2 = _g2;
        y1 = _y1;
        y2 = _y2;
    }

    function OtdsSign(uint256 c1, uint256 c2, 
        uint256 r1, uint256 r2, uint256 hmessage)
        public
    {
        require(state == 0, "Already signed.");
        // proof is (c1, c2, r1, r2)
        Curve.G1Point memory t1p = Curve.g1add(
                Curve.g1mul(y1, c1 % Curve.N()), 
                Curve.g1mul(g1, r1 % Curve.N()));
        Curve.G1Point memory t2p = Curve.g1add(
                Curve.g1mul(y2, c2 % Curve.N()), 
                Curve.g1mul(g2, r2 % Curve.N()));
        uint256 cp = uint256(keccak256(abi.encodePacked(
                g1.X, g1.Y, y1.X, y1.Y, 
                g2.X, g2.Y, y2.X, y2.Y, 
                t1p.X, t1p.Y, t2p.X, t2p.Y, hmessage)
                )) % Curve.N();
        require(addmod(c1, c2, Curve.N()) == cp, 
                "ERROR: Invalid proof.");
        // accept signature
        state = hmessage;
    }
}
\end{lstlisting}

  Ethereum distinguishes two types of costs related to smart contracts.
  On the one hand, \emph{transaction costs} are based on the costs for sending a smart contract to the blockchain, and depends on fixed costs for transactions and smart contracts, as well as the size of the smart contract to be deployed.
  On the other hand, \emph{execution costs} are based on the actual computations which need to be performed as the result of a transaction.
  This \emph{gas} is caluclated in \emph{gwei}, where $1 \text{ETH} = 10^9 \text{gwei}$;
  simple transactions require $21k$ gas, whereas complex transactions can easily exceed $1M$ gas.
  Therefore, for contracts to be practical they have not only to be implementable, but also with reasonable cost for the users.
  
  Table~\ref{tab:prices} shows the transaction costs and execution costs for the contracts presented above.
  For these contracts, the user needs to pay the transaction costs, while the delegate would need to pay for the execution costs.
  
  \begin{table}[h!]
    \begin{tabular}{c|c|c}
                         & Transaction   & Execution \\
                         & costs in kGas & costs in kGas \\
      \hline
      Plain EC multiplication     & 30    &  8 \\
      Basic scheme constructor     & 500   & 34 \\
      Basic scheme signature       & 67    & 41 \\
      Advanced scheme constructor  & 843   & 62 \\
      Advanced scheme signature    & 95    & 69 \\
    \end{tabular}
    \caption{Overview of costs measured (rounded to kGas).}
    \label{tab:prices}
  \end{table}

  At the time of writing this paper, $1 \text{ETH} \approx 2'500 \text{USD}$,\footnote{\url{https://coinmarketcap.com/currencies/ethereum/}} resulting in about $0.25\text{c}/\text{kGas}$.
  Thus, for instance, the transaction costs for a basic signature are about $16\text{c}$, while the execution costs are about $10\text{c}$, which can be considered practical for many sensitive applications compared to the costs caused by potential abuse of delegated rights. 
  It can be seen that the main costs are due when initializing the smart contract.
  These costs could easily be amortized by modifying the contract in a way that it can be called by many users and delegates, instead of using one contract per delegation.

\section{Conclusion}
  In this short paper we presented an alternative approach to enforcing the limited use of delegated cryptographic rights.
  Instead of relying on special-purpose hardware or aiming at disincentivizing delegates to abuse their rights, our approach leverages smart contracts to upper bound the number of invocations of delegated rights.
  We provided concrete implementations of the corresponding smart contracts for the Ethereum blockchain, proving the real-world applicabiltiy of our schemes.
  
  Future work will aim at extending the approach to additional applications beyond basic signature schemes.

\ifanonymous
\else
	\begin{acks}
The projects leading to this work have received funding from the European Union's Horizon 2020 research and innovation programme under grant agreement No 830929 (``CyberSec4Europe''), from the SESAR Joint Undertaking under grant agreement No 890456 (``SlotMachine''), and from the Austrian Research Promotion Agency (``FlexProd'').
	\end{acks}
\fi

\bibliographystyle{ACM-Reference-Format}
 \balance
\bibliography{cryptobib/crypto,cryptobib/abbrev0,additionalbib}

\end{document}